\documentclass[12pt]{article}
\usepackage{paper2e}
\usepackage{mydefs2e}

\begin{document}
\begin{titlepage}

\preprint{UMD-PP-00-68\\HUTP-00/A020}
\title{Electroweak Symmetry Breaking by Strong\\\medskip
Supersymmetric Dynamics at the TeV Scale}

\author{Markus A. Luty}

\address{Department of Physics,
University of Maryland\\
College Park, Maryland 20742, USA\\
{\tt mluty@physics.umd.edu}}

\author{John Terning, Aaron K. Grant}
\address{Department of Physics, Harvard University\\
Cambridge, Massachusetts 02138, USA\\
{\tt terning@schwinger.harvard.edu, grant@schwinger.harvard.edu}}

\begin{abstract}
We construct models in which electroweak symmetry is spontaneously
broken by supersymmetric strong dynamics at the TeV scale.
The order parameter is a composite of scalars, and the longitudinal
components of the $W$ and $Z$ are strongly-coupled bound states of scalars.
The usual phenomenological problems of dynamical electroweak symmetry
breaking are absent:
the sign of the $S$ parameter unconstrained in strongly interacting
SUSY theories, and
fermion masses are generated without flavor-changing neutral
currents or large corrections to the $\rho$ parameter.
The lightest neutral Higgs scalar can be heavier than $M_Z$
without radiative corrections from standard-model fields.
All the mass scales in the model can be naturally related in
low-scale models of supersymmetry breaking.
The $\mu$ problem can also be solved naturally, and the model can
incorporate perturbative unification of standard-model gauge
couplings with intermediate thresholds.
\end{abstract}

\date{June 19, 2000}

\end{titlepage}

\section{Introduction}
Understanding the origin of electroweak symmetry breaking is without
question the most important open problem in particle physics.
On the experimental side, despite a wealth of precision data that
shows convincingly that the electroweak interactions are described by a
spontaneously broken $SU(2)_W \times U(1)_Y$ gauge theory, we still have
no direct information about the
dynamics of electroweak symmetry breaking.
On the theoretical side, there are only a handful of mechanisms known
for electroweak symmetry breaking that can naturally explain the
enormous hierarchy between the weak scale $M_W \sim 100\GeV$ and more
fundamental scales such as the unification scale
$M_{\rm GUT} \sim 10^{16}\GeV$ and the Planck scale
$M_{\rm Planck} \sim 10^{18}\GeV$.
The oldest idea is that new QCD-like strong dynamics near the weak scale
are responsible for electroweak symmetry breaking \cite{TC}.
This idea, known as `technicolor', is currently
out of favor because of phenomenological problems and the
difficulty of constructing compelling models.
Perhaps the most attractive and well-studied idea is supersymmetry
(SUSY) \cite{SUSY}.
Most recently, there has been a great deal of interest in the idea
that the fundamental Planck scale is near the weak scale, thus obviating
the hierarchy problem.
In such scenarios the observed weakness of gravity compared to the weak
interactions is explained by the presence of large extra dimensions
felt only by gravity \cite{extradim} or by the effects of gravitational
curvature in extra dimensions \cite{warp}.

In this paper, we consider a new class of models in which electroweak
symmetry is broken by strong \emph{supersymmetric} dynamics at the
TeV scale.
Supersymmetry is assumed to be broken softly at the weak scale, but
this breaking is small enough to be viewed as a perturbation on the strong
dynamics.
Electroweak symmetry is broken by a VEV
for a composite operator made of scalars arising from a
non-perturbative `deformed moduli space' \cite{deform}.
The Nambu-Goldstone bosons that become the longitudinal components of
the $W$ and $Z$ are composites of scalars.
In this sense, the mechanism can be viewed as the `superpartner' of the
technicolor mechanism, in which the condensate and the longitudinal
component of the gauge bosons are fermion composites.
We therefore call this mechanism `$S$-color', where the `$S$' stands for
`super' or `scalar'.
We will show that these models elegantly avoid all of the problems of
technicolor models, and compare favorably with other SUSY models
in terms of naturalness and simplicity.
Most importantly, these models give a viable and well-motivated scenario for
strongly-coupled supersymmetric physics at the TeV scale.
The models have many interesting signatures, including a non-minimal
Higgs sector, non-standard Yukawa couplings, and an approximately
supersymmetric spectrum of strong resonances in the TeV region.

It is interesting to compare these models with non-supersymmetric
technicolor models.
Technicolor models have difficulty generating fermion masses without
generating large flavor-changing neutral currents \cite{TCFCNC}.
The models we consider have no problem with fermion masses because
they contain an elementary Higgs multiplet that gets a VEV by mixing
with the composite fields of the $S$-color sector.
The fermion masses therefore arise from ordinary Yukawa couplings, and
the usual GIM mechanism suppresses FCNC's.%
\footnote{Technicolor models with a GIM mechanism \cite{TCGIM}
or ``walking'' \cite{walking} can be constructed,
but the models are very complicated and require
nontrivial dynamical assumptions.}
Also, technicolor models generally give rise to large positive
contributions to the electroweak $S$ parameter from strong resonances in
the TeV region \cite{STC}.
In the models we consider, the sign of $S$ is not determined by any
currently known method.
Other radiative corrections are also naturally under control.

Compared to more traditional SUSY models, these models also have
a number of attractive features.
For example, the $\mu$ problem can be solved by the $S$-color dynamics.
Also, the lightest neutral Higgs boson can be significantly more massive than
$M_Z$ due to mixing with the composite states.
Perhaps the least appealing feature of these models is that the SUSY
breaking masses must be close to the $S$-color scale, even though they
do not originate from the $S$-color dynamics.
We will show that this can be natural if the $S$-color group is near a
conformal fixed point and is driven away from the fixed point by
low-scale SUSY breaking, for example from gauge-mediated SUSY
breaking.
We will present a model that incorporates this mechanism, together
with a dynamical solution to the $\mu$ problem and gauge
unification, all without excessive complication.

A model very similar to the ones considered here were discussed in \Ref{CK},
which appeared while this paper was in progress.
However, the model of \Ref{CK} has a massless fermion with
couplings to the $Z$, and is therefore ruled out.%
\footnote{The massless fermion can be avoided in the model of
\Ref{CK} by assuming a different structure
for the VEV's and introducing additional $B$-type soft masses.
See Section 2.}
Also, electroweak radiative corrections are not discussed in \Ref{CK}.
However, the idea that there can be strong approximately
supersymmetric dynamics at the TeV scale appears for the first time
in \Ref{CK}.

The models presented here also have some similarities with `bosonic
technicolor' models \cite{BTC}, which involve both SUSY and strong
dynamics near the TeV scale.
However, in bosonic technicolor, SUSY breaking scalar masses are large
compared to the strong dynamical scale, so the dynamics that breaks
electroweak symmetry is completely non-supersymmetric.
Therefore, in bosonic technicolor
the spectrum of strong resonances at the TeV scale is
non-supersymmetric, and the $S$ parameter is unsuppressed and positive.

This paper is organized as follows.
In Section 2, we analyze a minimal (but realistic) model that illustrates
the main features of the idea.
In Section 3, we consider an extension of the model analyzed in Section 2
that incorporates a solution to the $\mu$ problem.
In Section 4, we estimate the electroweak radiative corrections in this
class of models.  In Section 5 , we discuss a mechanism that can
explain the coincidence of the $S$-color scale and the scale of soft
SUSY breaking, and account for  gauge coupling unification.
Section 6 contains some speculations on phenomenology and our conclusions.
\section{A Minimal Model}
We now present a simple model that illustrates the main features of
the mechanism.
The non-Abelian symmetries of the model are
\beq
SU(2)_{\rm SC} \times SU(2)_L \times SU(2)_R,
\eeq
where $SU(2)_{\rm SC}$ is the $S$-color gauge group,
$SU(2)_L$ is the weak gauge group, and we only gauge the
$U(1)_Y$ subgroup of $SU(2)_R$ which is generated
by the $\tau_3$ generator.
The fields are
\beq
T_L \sim (\fund, \fund, \one),
\quad
T_R \sim (\fund, \one, \fund),
\quad
H \sim (\one, \fund, \fund),
\eeq
and two singlets $S_L, S_R$.
The field $H$ therefore contains a pair of doublet Higgs fields, and
the fields $T_L$ and $T_R$ are a supersymmetric version of minimal
technicolor \cite{TC}.
The theory has a tree-level superpotential
\beq\eql{Wtree}
W = \la_L S_L T_{L} T_{L} + \la_R S_R T_{R} T_{R}
+ \la_{H} H T_L T_R + \sfrac{1}{2} \mu H H.
\eeq
These terms break all global $U(1)$ symmetries, which is important for
avoiding massless fermions or axions. 
The gauge symmetries allow the addition of a superpotential term
\beq\eql{iso}
\De W = \la_H' H T_L (T_R \tau_3)
\eeq
that violates custodial $SU(2)$.
We will ignore this term for simplicity when discussing the effective
potential, but we will return to it when we discuss electroweak radiative
corrections.
The elementary Higgs fields, $H$,
are also assumed to have Yukawa couplings to
the quark and lepton fields.
These are important for generating the quark and lepton masses, but they
do not play a role for the vacuum structure as long as the squark and slepton
fields do not get VEV's.

The strong $S$-color dynamics has a global $SU(4)$ symmetry that is
broken only by standard-model gauge interactions and trilinear
superpotential couplings.
Under $SU(4)$, the $S$-colored fields transform as a fundamental:
\beq
T^j = \pmatrix{T_L \cr T_R \cr} \sim \fund,
\qquad
j = 1, \ldots, 4.
\eeq
The $SU(2)_{\rm SC}$ group has a deformed moduli space \cite{deform}.
This means that below the scale $\La$ where the theory
becomes strong, the light degrees of freedom correspond to the
`meson' fields $M^{jk} \propto T^j T^k$:
\beq
M^{jk} = -M^{kj} = \pmatrix{B_L \ep & \Pi \cr -\Pi^T & B_R \ep \cr},
\qquad
\ep \equiv \pmatrix{0 & 1 \cr -1 & 0 \cr},
\eeq
subject to the constraint
\beq\eql{qconstr}
\Pf(M) = B_L B_R - \det(\Pi) \ne 0.
\eeq
Under $SU(2)_L \times SU(2)_R$ the composite fields transform as
\beq
\Pi \sim (\fund, \fund),
\quad
B_L \sim (\one, \one),
\quad
B_R \sim (\one, \one).
\eeq



In order to be realistic, this theory must incorporate soft SUSY
breaking.%
\footnote{It is amusing to note that if we omit the $\mu$ term in
\Eq{Wtree}, then this model dynamically breaks SUSY \cite{IYIT}.
However, this cannot be the only source of SUSY breaking since it gives
very small (gauge-mediated) masses to standard-model gauginos and
scalars.
Even if we add soft SUSY breaking by hand, the model without the $\mu$
term gives rise to an `extra' Goldstino that couples to the $Z$.
Therefore, we must complicate the model to ensure that it does
\emph{not} dynamically break SUSY!}
%
Since the strong dynamics is responsible for breaking electroweak
symmetry, the required soft SUSY breaking terms are not much smaller
than the dynamical scale of the $S$-color dynamics.
However, we will see that \naive dimensional analysis (NDA) \cite{NDA,SUSYNDA}
indicates that it is sensible to treat soft SUSY breaking as a perturbation,

Denote the scale where the $S$-color dynamics becomes strong by $\La$.
In a normalization where the composite fields have kinetic terms of
order 1, the quantum constraint can be written \cite{SUSYNDA}
\beq\eql{NDAqconstr}
\det(\Pi) - B_L B_R = \sfrac{1}{2} f^2,
\eeq
and the effective superpotential is
\beq
W_{\rm eff} = f \left[
\la_L S_L B_L + \la_R S_R B_R + \la_H H \Pi \right]
+ \sfrac{1}{2} \mu H H ,
\eeq
where $f = \La / 4\pi$.
We have used our freedom to normalize the fields to set various coefficients
to 1;
in this normalization, all of the unknown strong interaction coefficients
appear in the effective \Kahler potential for the composite fields.

$SU(4)$ symmetry and NDA tells us that the effective \Kahler potential is
\beq
K_{\rm eff} &= f^2 k\left( \frac{\tr(M^\dagger M)}{2 f^2} \right)
\nonumber\\
\eql{Keff}
&= f^2 k\left( \frac{\Pi_0^\dagger \Pi_0^{\vphantom\dagger}
+ \Pi_A^\dagger \Pi_A^{\vphantom\dagger}
+ B_L^\dagger B_L^{\vphantom\dagger}
+ B_R^\dagger B_R^{\vphantom\dagger}}{f^2} \right),
\eeq
where $k$ is an unknown order-1 function.
We know that $k' > 0$ for all field values in order that the theory
has a positive kinetic term in the SUSY limit.

       From the above, we see that we require $f \sim 100\GeV$, which implies
$\La \sim 1\TeV$.
The soft masses must be of order $m_{\rm soft} \sim 100\GeV$, so NDA
implies that SUSY breaking perturbations are suppressed by
$m_{\rm soft} / \La \sim 1/4\pi$.
Some of our results rely on NDA, so it is reassuring to
note that NDA for soft SUSY breaking is known to work well in
supersymmetric theories where exact results are available \cite{LR}.
Note also that in QCD, the strange quark mass breaks $SU(3)$ flavor
symmetry by an amount $m_{\rm strange} / \La_{\rm QCD} \sim (100
\MeV) / (1\GeV)$,
a perturbation just as large as the one we are contemplating.
The fact that $SU(3)$ is a useful approximate symmetry in QCD is thus
further support that the expansion we are performing is sensible.

At this point, there is no explanation for the near coincidence of the
scales $f$ and $m_{\rm soft}$.
Also, the $\mu$ term must be put in by hand, and must be the same
order as $m_{\rm soft}$.
In Section 3 we will discuss extensions of this model that can address these
issues.
However, the present model gives a simple and realistic illustration of
the mechanism we are proposing. 

To solve the quantum constraint, we write
\beq
\Pi^j{}_k = \frac{1}{\sqrt{2}} (\Pi_0 \one_2 + i\Pi_A \tau_A)^j{}_k,
\quad
H^j{}_k = \frac{1}{\sqrt{2}} (H_0 \one_2 + i H_A \tau_A)^j{}_k,
\quad
\eeq
where $\tau_A$ ($A = 1,2,3$) are the Pauli matrices.
This gives
\beq
H \Pi = H_0 \Pi_0 + H_A \Pi_A,
\quad
\det(\Pi) = \sfrac{1}{2} \left( \Pi_0^2 + \Pi_A \Pi_A \right),
\eeq
{\it etc\/}.
Solving \Eq{NDAqconstr} for $\Pi_0$ gives
\beq\eql{P0solve}
\Pi_0 =
\left( f^2 + 2 B_L B_R - \Pi_A \Pi_A \right)^{1/2}.
\eeq
We therefore parameterize the moduli space by $B_L$, $B_R$, and $\Pi_A$;
this parameterization is non-singular for all vacua where $\avg{\Pi_0} \ne 0$.
In this way we obtain the unconstrained effective superpotential
\beq\eql{Weffsolve}
\bal
W_{\rm eff} = &f \Bigl\{
\la_L S_L B_L + \la_R S_R B_R
+ \la_H \left[ H_0 \Pi_0
+ H_A \Pi_A \right] \Bigr\}
\\
&+ \sfrac{1}{2} \mu (H_0^2 + H_A H_A),
\eal\eeq
where $\Pi_0$ is eliminated using \Eq{P0solve}.
Similarly,
$\Pi_0$ should also be eliminated in the effective \Kahler potential
\Eq{Keff}.


We now discuss the vacua in the SUSY limit.
The $H_0$, $H_A$, and $\Pi_A$ equations of motion give respectively
\beq
H_0 &= -{{\lambda_H f}\over{\mu}} \Pi_0,
\eql{H0EQM}
\\
f \lambda_H \Pi_A &= - \mu H_A,
\eql{HAEQM}
\\
H_0 \Pi_A &= H_A \Pi_0,
\eql{PAEQM}
\eeq
where $\Pi_0$ is given by \Eq{P0solve}.
Substituting \Eq{HAEQM} into \Eq{PAEQM} reproduces \Eq{H0EQM}, so we
find three flat directions.
The moduli space of vacua includes a subspace  where
$SU(2)_L \times SU(2)_R \rightarrow SU(2)$ i.e. the gauge symmetry
breaks as $SU(2)_L \times U(1)_Y \rightarrow U(1)_{em}$.
In these vacua, electroweak symmetry is broken in the correct pattern in
the SUSY limit, and the three flat directions are associated with the
Nambu-Goldstone bosons of the symmetry breaking.

To obtain a realistic model we must include soft SUSY breaking with
$m_{\rm soft} \sim \La / 4\pi$.
We must then check that there are choices for the fundamental soft masses
where electroweak symmetry is broken.
An important point is that the potential has no global $U(1)$
symmetries, so there is no danger of obtaining a weak-scale axion.
The potential is not calculable in this model, because when we include
soft SUSY breaking the potential depends on the full functional form of
the effective \Kahler potential, parameterized by the function $k$ defined
in \Eq{Keff}.
This is simply because in units where $f = 1$, the \Kahler potential is an
order-1 function of an order-1 argument.
Derivatives of the \Kahler potential appear multiplicatively in the
potential, and do not affect the VEV's in the SUSY limit.
However, for $m_{\rm soft} \sim f$,
these multiplicative corrections are parametrically as important as the
soft mass contributions to the potential.
%
Without knowledge of the \Kahler potential we cannot determine rigorously
whether vacua of the desired form exist.
However, given the large number of free parameters in the soft masses,
it is reasonable to assume that there are vacua that break
electroweak symmetry in the desired fashion.%
\footnote{In the limit where the superpartners decouple this model becomes
minimal technicolor.
In minimal technicolor, the vacuum aligns to break electromagnetism,
due to effects of standard-model gauge loops.
In the present model, the superpotential couplings as well as standard-model
gauge couplings break the accidental global symmetries of the strong dynamics.
There is no reason to think that the problems of minimal technicolor are
present in our model.}

We now turn to the fermion masses.
We first neglect gaugino masses.
In the presence of a nontrivial \Kahler potential and SUSY breaking,
the fermion mass matrix is proportional to
\beq
m_{ab} = \avg{W_{ab}} + \avg{K_{ab}{}^c} \avg{F^\dagger_c},
\eeq
where we denote the fields by $\Phi^a$ and
$W_a = \partial W / \partial \Phi^a$,
$W^a = \partial W / \partial \Phi^\dagger_a$, {\it etc\/}.
The physical fermion mass matrices are given by matrix products of
$m_{ab}$ and $\avg{K^a{}_b}$, but the important fact for our purposes
is that massless fermions are present if and only if $\det(m) = 0$.%
\footnote{This assumes that the \Kahler potential is nonsingular.}
The determinant of $m$ is thus an important diagnostic,
and we find that it is nonzero for general VEV's.
The precise expression depends on the form of the effective \Kahler
potential, and is complicated and unenlightening;
for example, if we assume that the \Kahler potential is
$\tr(M^\dagger M)$, and assume
$\avg{\Pi_A} = 0$, $\avg{B_L} = \avg{B_R} = 0$, we find
\beq\eql{mfermdet}
\det(m) = -\mu f^7 \la_H^3 \la_L^2 \la_R^2 (\mu \avg{H_0} + \la_H f)^3.
\eeq
%
%
This vanishes for SUSY vacua (see \Eq{H0EQM}), but is
nonzero (and nonsingular) for general VEV's.
When we include the gauginos, there are mass terms that mix the gauginos
with some of the fermions above, as part of the SUSY Higgs effect.
These mixing mass terms are of order $M_Z$, so the
non-vanishing of the determinant above shows that there are no
massless fermions
in the limit $g_{1,2} \to 0$. 
This is important because it shows that there are no fermions whose
mass comes entirely from the SUSY Higgs effect, so there is no reason that all
fermions cannot be heavier than $M_Z$.
The nonzero electroweak gauge couplings can in principle give rise to light
fermions, but only for special parameter choices.
We conclude that the fermion masses do not present a phenomenological
problem for this model.%
%

%
%
%
%

An undesirable feature of the present model is that it contains an
explicit `$\mu$ term'.
This term must be of order the weak scale:
if $\mu$ is large compared to the weak scale,
the elementary Higgs fields $H$ decouple and we do not generate
quark and lepton masses;
if $\mu$ is too small we have light fermions (see \Eq{mfermdet}).
(The only difference from the $\mu$ problem in the MSSM is that the
present model can break electroweak symmetry for any value of $\mu$.)
In the next Section, we show that
a simple modification of the model can solve this problem.

We close this Section by considering what happens when $\la_L = \la_R = 0$.
Then the singlets $S_L, S_R$ decouple, and we obtain the model of
\Ref{CK}.
This model has an anomaly-free global $U(1)_{SB} \times U(1)_R$
symmetry, where the $U(1)_{SB}$ is `$S$-baryon' number.
The $U(1)_R$ symmetry is broken explicitly by soft SUSY breaking terms.
\Ref{CK} assumes that $\avg{B_L} = \avg{B_R} = 0$ in order to avoid
spontaneously breaking the $U(1)_{SB}$.
In this case, we find that there is a massless `baryon' fermion in the
spectrum.
This fermion has unsuppressed
couplings to the $Z$, and is therefore ruled out.
The massless fermion can be avoided by allowing
$\avg{B_L}, \avg{B_R} \ne 0$.
In order to avoid a weak-scale axion, $U(1)_{SB}$ must be broken explicitly.
This can be done by adding the $B$-type soft SUSY breaking
mass terms $T_L T_L + \hc$,
$T_R T_R + \hc$ to the potential.
The origin of these terms in a specific model of SUSY breaking may be
difficult to understand, since there are no terms with the same symmetries
in the SUSY part of the theory.
\section{An Improved Model}
We can eliminate the $\mu$ problem simply by replacing the
$\mu$ term with a cubic interaction\footnote{This
is similar in spirit to the next-to-minimal supersymmetric standard
model \cite{NMSSM}}:
\beq\eql{Wimproved}
W = \la_L S_L T_{L} T_{L} + \la_R S_R T_{R} T_{R}
+ \la_{H} H T_L T_R + \sfrac{1}{2} y (S_L + S_R) H H.
\eeq
The symmetry between the $S_L$ and $S_R$ cubic couplings is
not essential; it merely simplifies the form of the
VEV's in the model.
We can also include further cubic interactions
for the singlets $S_L$ and $S_R$, but these do not
lead to qualitatively different results.
Note that all global $U(1)$ symmetries are broken.

In the SUSY limit, the VEV's are determined by
\beq
{{\partial W}\over{\partial S_L}} =& f \lambda_L B_L
+ {{y}\over{2}}(H_0 H_0 + H_A H_A),
\nonumber\\
{{\partial W}\over{\partial S_R}} =& f \lambda_R B_R
+ {{y}\over{2}}(H_0 H_0 + H_A H_A),
\nonumber\\
{{\partial W}\over{\partial B_L}} =& f \lambda_L S_L
- {{f \lambda_H H_0 B_R}\over{\Pi_0}},
\nonumber\\
{{\partial W}\over{\partial B_R}} =& f \lambda_R S_R
- {{f \lambda_H H_0 B_L}\over{\Pi_0}},
\\
{{\partial W}\over{\partial H_0}} =&
f \lambda_H \Pi_0 + y (S_L+S_R) H_0,
\nonumber\\
{{\partial W}\over{\partial H_A}} =&
f \lambda_H \Pi_A + y (S_L+S_R) H_A,
\nonumber\\
{{\partial W}\over{\partial \Pi_A}} =&
f \lambda_H \left( H_A -
{{H_0 \Pi_A}\over{\Pi_0}}\right).
\nonumber
\eeq
Using the $H_0$ equation we can see that the $\Pi_A$ and $H_A$ equations
are equivalent, so we again find three flat directions.
Solving the remaining equations for the special case $\avg{H_A} = 0$,
we obtain
\beq
\bal
\avg{H_0} &= \left( {{2 \lambda_L \lambda_R}\over{9 y^2}}\right)^{1/4} f, \\
\avg{S_L} = \avg{S_R} &= \pm \lambda_H
\left( {{2 }\over{9 y^2\lambda_L \lambda_R}}\right)^{1/4} f, \\
\avg{B_L} &= -\left(
{{ \lambda_R }\over{18 y^2\lambda_L }}\right)^{1/2} f, \\
\avg{B_R} &= -\left(
{{ \lambda_L }\over{18 y^2\lambda_R }}\right)^{1/2} f. \\
\eal
\eeq
We see that there are points on the moduli space where electroweak
symmetry is broken in the correct pattern.
In addition, the nonzero VEV for the singlets gives an effective
$\mu$ term for the Higgs doublets.

The inclusion of soft SUSY breaking proceeds as for the simpler model
above.  
The details are not enlightening, and will not be given here.
We expect that there is a vacuum with the desired properties
for reasonable choices of soft masses.

We also computed the determinant of the fermion mass matrix to check
that there are no light fermions.
As in the previous model,
we find that the fermion determinant is nonzero and unsuppressed
for non-supersymmetric VEV's.
The discussion is similar to that for the simpler model, but the
expressions are more complicated.

\section{Electroweak Radiative Corrections}
We now discuss electroweak radiative corrections in these models.
This is particularly interesting because it is generally believed that
models in which electroweak symmetry is broken by strong dynamics are
strongly constrained by the electroweak $S$ parameter \cite{STC}.
However, we show that in the present model this is not the case.
We will show that other radiative corrections are also small.

\subsection{The $S$ Parameter}
The $S$ parameter can be viewed as a gauge kinetic mixing term between
$SU(2)_L$ and $U(1)_Y$ gauge groups in an effective theory below the
scale $\La$ where electroweak symmetry is broken:
\beq\eql{Sdef}
{\cal L}_{\rm eff} = -{{S}\over{16 \pi}} g g' F^{\mu \nu}_L F_{\mu\nu Y}~.
\eeq
In our model, the leading contribution to $S$ comes from operators of the form
\beq
\frac{c}{\La^2}
\myint d^4\th\, \tr \left[ {\overline \nabla}\nabla
(M^\dagger e^V) {\overline \nabla}\nabla M \right] + \hc,
\eeq
where $\nabla_\alpha$ is the gauge covariant SUSY derivative, and $c \sim 1$
by NDA.
Putting in VEV's for $M$ we find
\beq\eql{KS}
S \sim \frac{c}{\pi} \sim \pm 0.3.
\eeq
Note that there are no large multiplicity factors, since there is only a
single electroweak doublet charged under a strongly coupled $SU(2)$ gauge
group.

The most recent Particle Data Group analysis of precision electroweak
data gives $S = -0.16 \pm 0.14$ for the standard model with $m_{h^0} = M_Z$,
and $S = -0.26 \pm 0.14$ for $m_{h^0} = 300\GeV$ \cite{PDG}.
We see that the estimates above are compatible with the data, provided that
the sign of $S$ can be negative.

It is therefore very encouraging that the sign of the contribution \Eq{KS}
to $S$ is not determined by the usual arguments in QCD-like theories,
because of the crucial role played by scalars in these theories.  Clearly
we cannot use QCD data to directly estimate $S$ in these theories.  An
alternative approach in QCD-like theories uses the Weinberg sum rules
\cite{WSR} together with the less rigorous resonance saturation
``aproximation'' to estimate $S$.  The first step is to use the operator
product expansion (OPE) to find the short-distance behavior of the
current-current correlation function relevant for $S$:
\beq
\bal \myint d^4
x\, e^{i p \cdot x} & \bra{0} T J_{L A}^\mu(x) J_{R B}^\nu(0) \ket{0} \\
&\sim \frac{g^{\mu\nu} - p^\mu p^\nu/p^2}{p^2} \bra{0} (T_L^\dagger \tau_{L
A} T_L) (T_R^\dagger \tau_{R B} T_R) \ket{0} + \cdots, \eal
\label{correlator}
\eeq
where $T_{L,R}$ are the scalar components of the $S$-colored fields.  In
QCD,
the
leading operator on the \rhs is a quartic fermion term, and the correlation
function behaves as $1/p^4$ rather than $1/p^2$.  This implies that only
the first Weinberg sum rule holds in the present class of theories.  The
Weinberg sum rules can be written in terms of spectral density functions
for vector and axial-vector channels \cite{WSR}.
Making the assumption that
these sum rules are approximately
saturated by the lowest-lying single particle intermediate states with
vector and axial-vector quantum numbers, the first and second sum rules yield
\beq
f_\rho^2-f_A^2 &= f_\pi^2
\eql{wein1}\\
f_\rho^2 m_\rho^2-f_A^2 m_A^2 &= 0~.
\eql{wein2}
\eeq
With these assumptions, $S$ is given by \cite{Ssat}
\beq
S = 4 \pi\left(
{{f_\rho^2}\over{m_\rho^2}}-{{f_A^2}\over{m_A^2}}\right)~.
\eeq
Using both sum rules \Eqs{wein1} and \eq{wein2}, one one can show 
that $m_\rho <
m_A$, $f_A < f_\rho$ and hence that $S>0$.
However in a SUSY theory with only the first Weinberg sum rule 
\Eq{wein1} we can
reach no conclusion as to the sign of $S$.
Abandoning the saturation approximation we have even less information.

Yet another  approach to determining the sign of $S$ is to apply
Vafa-Witten \cite{VafaWitten} positivity arguments to current-current
correlators.  However, in spite of significant efforts along these lines,
the sign of $S$ has not been determined in this way for a QCD-like theory
\cite{Nussinov}.  Furthermore, the Vafa-Witten arguments
generally break down in theories (such as supersymmetric theories)
that include scalars with Yukawa couplings.
We conclude that there is no reason to believe that $S$ cannot be
negative in this class of models.

In the remainder of this Section, we will argue that the $F$ term
contributions to $S$ are much smaller than the contribution of
\Eq{KS}.
This is a somewhat surprising result, because there is an operator
\beq\eql{Aaron}
\frac{1}{\La^2}
\myint d^2\th\, (W^\al)^j{}_k (W_\al)^\ell{}_n \epsilon_{j\ell rs}
   M^{kn} M^{rs} + \hc
\eeq
that is invariant under all symmetries, and that gives a nonzero value for $S$.
NDA implies that the the contribution to $S$ from this operator is the same
as \Eq{KS}, so the conclusions above would not be affected if this operator
is present.
The remainder of this Subsection is therefore primarily of
theoretical interest,
and the reader interested mainly in the results is urged to skip
to the next
Subsection.

We begin by classifying the possible $F$ terms that can contribute to $S$.
For this, it is important to keep track of the $SU(4)$ symmetry of the strong
dynamics.
The elementary Higgs fields and composite `meson' fields fall into
$SU(4)$ representations
\beq
\Si_{jk} = \pmatrix{\la_L S_L \ep & \la_H H \cr
-\la_H H^T & \la_R S_R \ep \cr}
\sim \overline{\raisebox{-4pt}{\asymm}}\,,
\qquad
M^{jk} = \pmatrix{B_L \ep & \Pi \cr -\Pi^T & B_R \ep \cr}
\sim \raisebox{-4pt}{\asymm}\,,
\eeq
while the $SU(2)_L \times SU(2)_R$ field strengths are in the $SU(4)$
adjoint:
\beq
(W_{\alpha})^j{}_k \sim \pmatrix{ W_{L\al} & 0 \cr 0 & W_{R\al} \cr}
\sim {\bf Ad}.
\eeq
In this notation, the tree-level superpotential of the model of Section
3 is
\beq
W = \Si_{jk} M^{jk} + y^{jk\ell n pq} \Si_{jk} \Si_{\ell n}\Si_{p q},
\eeq
where $y^{jk\ell mpq}$ contains the cubic term in \Eq{Wimproved}.
In addition, there is an anomaly-free $U(1)_R$ symmetry with
\beq
R(M) = 0,
\quad
R(\Si) = 2,
\quad
R(W_\alpha) = 1,
\quad
R(y) = -4.
\eeq
This $SU(4) \times U(1)_R$ symmetry strongly constrains the form of
operators that can appear in the effective theory.

We now show that the only operator allowed by these symmetries
that can contribute to $S$ at the level of \Eq{KS} is \Eq{Aaron}.
We first consider contributions that do not vanish in the SUSY limit.
In the SUSY limit, any operator that contributes to a gauge kinetic mixing
must have the form
\beq
\myint d^2\th\, (W^\al)^j{}_k (W_\al)^\ell{}_n
\scr{F}^{k n}{}_{j\ell}(M, \Si, y) + \hc
\eeq
We can expand $\scr{F}$ in a power series in $y$, and NDA tells us
that the terms
proportional to powers of $y$ are suppressed by powers of $1/4\pi$.
We therefore consider only terms independent of $y$.%
\footnote{There are contributions such as
$\scr{F}^{jk}{}_{\ell n} \sim y^{jkpqrs} \Si_{pq} \Si_{rs} (M^{-1})_{\ell n}$
that are allowed by all symmetries.}
$\scr{F}$ cannot depend on $\Si^{-1}$ because it must have a smooth
limit as $\Si \to 0$.
($\scr{F}$ can depend on both $M$ and $M^{-1}$ to construct invariants since
$\Pf(M) \ne 0$ ensures that $M$ is invertible.)
Therefore $U(1)_R$ invariance then does not allow any dependence of the
$y$-independent part of $\scr{F}$ on $\Si$.
We are left with terms where $\scr{F}$ is a function of $M$ only.
It is not hard to see that \Eq{Aaron} is the only possibility, taking
into account the quantum constraint.

However, we claim that the operator \Eq{Aaron} cannot be present in the theory
because it cannot arise from the theory with an additional massive flavor.
Consider an $SU(2)$ gauge theory with 6 fundamentals, and a mass term
$W = m_{jk} M^{jk}$ that gives mass to 2 fundamentals.
Near the origin of moduli space, this theory has a weakly-coupled `s-confined'
\cite{deform} description in terms of the unconstrained
meson fields $M^{jk}$ and
a dynamically generated superpotential
$W_{\rm dyn} \propto \Pf(M)$.
The theory has an anomaly-free $U(1)_R$ symmetry with $R(M) = \sfrac{2}{3}$,
$R(m) = \sfrac{4}{3}$.
This forbids all terms of the form
\beq
\myint d^2\th\, (W^\al)^j{}_k (W_\al)^\ell{}_n
\scr{F}^{k n}{}_{j \ell}(M, m) + \hc
\eeq
where $\scr{F}$ is nonsingular in the limit $m \to 0$.
For example, the term
\beq
\myint d^2\th\, (W^\al)^j{}_k (W_\al)^\ell{}_n \ep_{j\ell p q r s}
M^{k n} M^{p q} (m^{-1})^{r s}
\eeq
is invariant under all symmetries, and reduces to \Eq{Aaron} if the matrix
$m_{jk}$ has rank 2, but it has a singular limit as $m \to 0$.
Therefore, it cannot appear in the effective theory below the scale $\La_6$,
where the theory with 6 fundamentals becomes strong.

We now turn to the possibility that the operator \Eq{Aaron} is generated
in the theory with 6 fundamentals when we integrate out the massive modes.
For $m \ll \La_6$ we can integrate out the massive modes using
the confined description.
The operator \Eq{Aaron} does not appear at any order in the perturbative
expansion, as follows from conventional perturbative non-renormalization
theorems.
We believe that there are no non-perturbative corrections to the matching that
can give \Eq{Aaron}, since the effective theory is simply a weakly coupled
Wess-Zumino model.
This argument is valid only for $m \ll \La$, but there can be no phase
transitions as a function of a holomorphic coupling in the SUSY limit, so
we conclude
that the operator \Eq{Aaron} vanishes even for $m \gg \La$.
We can summarize this by saying that the operator \Eq{Aaron} in the theory with
4 fundamentals is not generated despite the fact that it is invariant under all
symmetries  we do not have a generic superpotential in the ultraviolet theory
(with 6 fundamentals).


There are additional contributions to the $S$ parameter from SUSY breaking
terms such as
\beq
\myint d^4 \th\, (W^{\al})^j{}_k (W_\al)^\ell{}_n (M^\dagger)_{j\ell}
(\Si^\dagger)^{k n} + \hc
\eeq
However, NDA shows that these are highly suppressed because the
VEV's of fields and SUSY breaking parameters are of order
$\La / 4\pi$, while the mass scale that suppresses such higher
dimension operators is $\La$.

\subsection{The $T$ Parameter}
We now turn to the isospin-breaking
$T$ parameter.
In an effective Lagrangian language, $T$ is proportional to the difference
between the $W_3$ and $W^\pm$ mass term obtained by integrating out the
$S$-color states at the TeV scale.
The only large contribution from the $S$-color dynamics comes from the
isospin breaking term \Eq{iso}, which gives a shift in the $Z$ mass
of order
\beq
\Delta M_Z^2 \sim {{ (\la_H' f)^2}\over{16 \pi^2}}.
\eeq
Comparing this to the shift induced by the top quark, we obtain
\beq
\frac{T_{\rm SC}}{T_{\rm top}}
\sim \frac{(\la_H' f)^2}{m_{\rm top}^2}.
\eeq
For $\la'_H \sim 1$, the $S$-color contribution is as large as the top
contribution, but we can easily obtain an acceptable contribution for
moderately small values of $\la'_H$.%
\footnote{Note that $f$ is somewhat smaller than $v = 246\GeV$, since
electroweak symmetry breaking is distributed between four Higgs doublets.}

\section{Further Model-building}
Why is $f \sim m_{\rm soft}$?
It could be a coincidence just like $f_\pi \sim m_{\rm strange}$.
However, it is not difficult to construct models where the strong interaction
scale of $S$-color is fixed by SUSY breaking.
The idea is that the $S$-color gauge dynamics is near a strongly coupled
fixed point, and is perturbed away from this fixed point by SUSY breaking,
similar to `postmodern' technicolor theories \cite{PMTC}.
For example, in our model, $SU(2)_{\rm SC}$ would have an infrared fixed point
if there were four or five flavors rather than two.
If the additional $S$-colored fields are electroweak singlets and receive
masses somewhat larger than $m_{\rm soft}$, then the $S$-color gauge coupling
rapidly becomes strongly coupled near the scale $m_{\rm soft}$.

For an explicit example of this type of model, consider a theory with
four $S$-color flavors, one  electroweak doublet and two electroweak
singlets, as well as a gauge singlet
$X$ with a superpotential term
\beq
\De W = \frac{\ka}{3} X^3.
\eeq
Suppose that $X$ gets a negative soft mass squared of order
$m_{\rm soft}^2 \sim (100\GeV)^2$.
(We will discuss how this can happen below.)
The potential for $X$ is
\beq
V \sim -m_{\rm soft}^2 |X|^2 + |\ka|^2 |X|^4,
\eeq
with a minimum at $\avg{X} \sim m_{\rm soft} / \kappa$.
For small $\kappa$, $X$ gets a VEV larger than $m_{\rm soft}$.
If $X$ has Yukawa couplings to the $S$-colored electroweak singlets,
then they can be integrated out and the $S$-color gauge coupling will
rapidly go from fixed point behavior to strong coupling.
This threshold is not supersymmetric, but
$\avg{F_X} / \avg{X} \sim m_{\rm soft}$.
This SUSY breaking feeds into $S$-colored superpartners through
$S$-color gauge mediation, but this gives corrections smaller than
$m_{\rm soft}$.

One might also wonder what becomes of unification since the $S$-color
particles we have added to the MSSM are not in complete $SU(5)$ multiplets.
There are many ways to achieve unification with the addition of
further particles.
An attractive possibility is that these may be responsible
for gauge-mediating SUSY breaking to the ordinary superpartners.
A simple example is to add a vector-like right-handed up quark to
the model of this Section.
With the $S$-color particles responsible for electroweak
symmetry breaking described above this makes an approximate $SU(5)$
multiplet.
The SM gauge couplings unify  if the vector-like quark mass is near $10^6$~GeV.
If the vector-like quark is the gauge messenger of SUSY breaking for the usual
superpartners then the spectrum of superpartner masses will differ from
the usual scenarios, since the messengers are not complete $SU(5)$ multiplets.

%
%

\section{Conclusions}
We have presented models that dynamically break electroweak symmetry
via strong supersymmetric (`$S$-color') dynamics.
Our analysis of the dynamics is based on exact non-perturbative results
in supersymmetric gauge theories, and is therefore on a firm theoretical
foundation.
The failure of the second Weinberg sum-rule
shows that the $S$ parameter can have either sign, and is thus potentially
consistent with precision electroweak data.
One of our models gives a solution to the `$\mu$ problem.'
Simple extensions of these models can relate the supersymmetry breaking scale
to the $S$-color scale and allow for gauge coupling unification.

The phenomenology of these models is very exciting.
The spectrum contains the MSSM spectrum with two extra (composite) Higgs
doublets.
In addition, the theory is strongly coupled  with a rich
spectrum of supersymmetric
strong interaction resonances in  the TeV range, and therefore
will exhibit anomalous $WW$ scattering.
Yukawa couplings are larger than in the MSSM (or the standard model),
since electroweak symmetry breaking is distributed between composite
and fundamental Higgses.
We hope that this approach to electroweak symmetry breaking will stimulate
further theoretical and experimental
work on the possibility of strong dynamics at the TeV scale.

\section*{Acknowledgments}
M.A.L. thanks N. Arkani-Hamed for encouragement, and
thanks the theoretical particle physics groups at Stanford
University, Lawrence Berkeley National Laboratory, and Harvard
University for hospitality during the course of this work.
We also thank H. Georgi for helpful discussions.
M.A.L. was supported by the National
Science Foundation under grant PHY-98-02551,
by the Alfred P. Sloan Foundation.  The work of A.G. and J.T. is supported
in part by the NSF under grant PHY-98-02709.


\end{document}